\documentclass[trackchanges,twocolumn]{aastex61}
\usepackage{ulem}

\newcommand{\lsim }{{\lower0.8ex\hbox{$\buildrel <\over\sim$}}}
\newcommand{\gsim }{{\lower0.8ex\hbox{$\buildrel >\over\sim$}}}

\def\apjl{ApJL}
\def\aap{A\&A}

\def\apjs{ApJ Supp}

\def\chandra{\emph{Chandra}}

\def\hst{${\it HST}$}
\def\acs{${\it HST/ACS}$}

\def \gaia{{\it Gaia}}

\def\ergs{erg s$^{-1}$}
\def\simge{\mathrel{%
  \rlap{\raise 0.511ex \hbox{$>$}}{\lower 0.511ex \hbox{$\sim$}}}}
\def\simle{\mathrel{
  \rlap{\raise 0.511ex \hbox{$<$}}{\lower 0.511ex \hbox{$\sim$}}}}
\newcommand{\Msun}{\ifmmode {M_{\odot}}\else${M_{\odot}}$\fi}
\newcommand{\Lsun}{\ifmmode {L_{\odot}}\else${L_{\odot}}$\fi}
\newcommand{\Rsun}{\ifmmode {R_{\odot}}\else${R_{\odot}}$\fi}

\shorttitle{TMSP candidate in Terzan 5}
\shortauthors{Bahramian et al.}

\begin{document}
\title{The MAVERIC survey: A transitional millisecond pulsar candidate in Terzan 5}

\correspondingauthor{Arash Bahramian}
\email{bahramian@pa.msu.edu}

\author[0000-0003-2506-6041]{Arash Bahramian}
\affil{Department of Physics and Astronomy, Michigan State University, East Lansing, Michigan 48824, USA}
\affil{International Centre for Radio Astronomy Research – Curtin University, GPO Box U1987, Perth, WA 6845, Australia}

\author[0000-0002-1468-9668]{Jay Strader}
\affil{Department of Physics and Astronomy, Michigan State University, East Lansing, Michigan 48824, USA}

\author[0000-0002-8400-3705]{Laura Chomiuk}
\affil{Department of Physics and Astronomy, Michigan State University, East Lansing, Michigan 48824, USA}

\author[0000-0003-3944-6109]{Craig O. Heinke}
\affil{Physics Department, University of Alberta, 4-183 CCIS, Edmonton, AB T6G 2E1, Canada}

\author[0000-0003-3124-2814]{James C. A. Miller-Jones}
\affil{International Centre for Radio Astronomy Research – Curtin University, GPO Box U1987, Perth, WA 6845, Australia}

\author{Nathalie Degenaar}
\affil{Anton Pannekoek Institute for Astronomy, University of Amsterdam, Science Park 904, NL-1098 XH, Amsterdam, the Netherlands}

\author[0000-0003-3906-4354]{Alexandra J. Tetarenko}
\affil{Physics Department, University of Alberta, 4-183 CCIS, Edmonton, AB T6G 2E1, Canada}

\author[0000-0003-4553-4607]{Vlad Tudor}
\affil{International Centre for Radio Astronomy Research – Curtin University, GPO Box U1987, Perth, WA 6845, Australia}

\author[0000-0002-4039-6703]{Evangelia Tremou}
\affil{AIM/CEA Paris-Saclay, Universit\'e Paris Diderot, CNRS, F-91191 Gif-sur-Yvette, France}
\affil{Department of Physics and Astronomy, Michigan State University, East Lansing, Michigan 48824, USA}

\author{Laura Shishkovsky}
\affil{Department of Physics and Astronomy, Michigan State University, East Lansing, Michigan 48824, USA}

\author[0000-0002-3516-2152]{Rudy Wijnands}
\affil{Anton Pannekoek Institute for Astronomy, University of Amsterdam, Science Park 904, NL-1098 XH, Amsterdam, the Netherlands}

\author[0000-0003-0976-4755]{Thomas J. Maccarone}
\affil{Department of Physics and Astronomy, Texas Tech University, Box 41051, Lubbock, TX 79409–1051, USA}

\author[0000-0001-6682-916X]{Gregory R. Sivakoff}
\affil{Physics Department, University of Alberta, 4-183 CCIS, Edmonton, AB T6G 2E1, Canada}

\author[0000-0001-5799-9714]{Scott Ransom}
\affil{National Radio Astronomy Observatory, Charlottesville, VA 22903, USA}

\begin{abstract}
Transitional millisecond pulsars are accreting millisecond pulsars that switch between accreting X-ray binary and millisecond radio pulsar states. Only a handful of these objects have been identified so far. Terzan 5 CX1 is a variable hard X-ray source in the globular cluster Terzan 5. In this paper, we identify a radio counterpart to CX1 in deep Very Large Array radio continuum data. \chandra\ observations over the last fourteen years indicate that CX1 shows two  brightness states: in 2003 and 2016 the source was the brightest X-ray source in the cluster (at L$_X \sim 10^{33}$ erg s$^{-1}$), while in many intermediate observations, its luminosity was almost an order of magnitude lower.
We analyze all available X-ray data of CX1, showing that the two states are consistent with the spectral and variability properties observed for the X-ray active and radio pulsar states of known transitional millisecond pulsars.  Finally, we discuss the prospects for the detection of CX1 as a radio pulsar in existing timing data.
\end{abstract}


\section{Introduction} \label{sec:intro}
\subsection{Transitional Millisecond Pulsars}
Millisecond pulsars (MSPs) are old radio pulsars that have been spun up through accretion from a companion \citep{Alpar82}. Neutron star low-mass X-ray binaries (LMXBs) are progenitors of millisecond pulsars. This has been confirmed by the detection of coherent millisecond X-ray pulsations in the LMXB SAX J1808.4--3658 \citep{Wijnands98}, and later in a dozen other LMXBs \citep{Patruno12b,Patruno17}. The evolutionary link connecting LMXBs and MSPs has recently been found in the form of transitional millisecond pulsars (tMSPs): neutron stars that move back and forth between accretion-powered LMXB and rotation-powered MSP states on timescales as short as weeks \citep[e.g.,][]{Archibald09,Papitto13}. This exciting discovery implies that there is not a simple one-way evolution from LMXBs to MSPs, but instead an extended evolutionary phase in which neutron stars switch between these two different states.

Three tMSPs have now been confirmed through multi-wavelength studies; PSR J1023+0038 \citep{Archibald09}, IGR J18245-2452 in the globular cluster M 28 \citep{Papitto13}, and XSS J12270-4859 \citep{Bassa14}. All three have shown transitions between a radio pulsar state and a low-level accretion state \citep[e.g.,][]{Stappers14,Papitto13,Bassa14,Roy15}. During the radio pulsar state, radio pulsations are often (but not always) detected, and the radio spectrum is a steep power-law ($S \propto \nu^{\alpha}$) with typical spectral index similar to other MSPs, $\alpha \sim -1.6$ to $-1.8$ \citep[e.g.,][]{kramer98}, while the X-ray emission is relatively faint \citep[$L_X \sim 10^{32}~{\rm erg~s}^{-1}$ ,][]{Bogdanov14,Linares14b,Roberts15}. In the low-level accretion state, radio pulsations cease, and bright radio emission with a flat radio spectrum ($\alpha \sim 0$) and rapid variability \citep[e.g.,][]{Deller15} is thought to be produced by a partially self-absorbed synchrotron jet, while the X-ray luminosity increases to $L_X \sim 10^{33-34}~{\rm erg~s}^{-1}$ \citep[e.g.,][]{Bogdanov15b}. During this state, X-ray pulsations  thought to be powered through accretion have been detected \citep{Papitto13, Archibald15, Papitto15, Jaodand16}. In both states the X-ray spectrum is relatively hard, with power-law photon index 1-1.5 ($\frac{dN}{dE}\propto E^{-\Gamma}$). Furthermore, in the low-level accretion state, tMSPs tend to show significant variability, and occasional flares \citep[e.g.,][]{Tendulkar14,Takata14,Papitto15}. They can also show two dominant ``modes'' of brightness (labeled active and passive modes), whose flux can differ by a factor of a few \citep[e.g.][]{Bogdanov14}.

It is worth noting that  radio emission from tMSPs during the low-level accretion state is thought to be due to a partially self-absorbed jet \citep{Deller15}. However, \citet{Bogdanov18} recently showed a sharp anti-correlation between radio and X-ray brightness in PSR J1023+0038 during this state. This is at odds with the standard accretion-jet coupling. The synchronous rise and decay of the radio emission during the drop into, and rise out of, the X-ray low mode suggests that the radio-emitting region is causally-connected to the jets, which implies it must be very compact (tens of light seconds). Furthermore, \citet{Ambrosino17} recently detected optical pulsations from PSR J1023+0038 during the low-level accretion state, which indicates that the optical (and perhaps X-ray) emission in this state from these systems may be powered through magnetospheric processes.

Correlated changes in the X-rays and radio, along with changes in the optical and gamma-rays (harder to study in distant, crowded globular clusters) have clearly identified X-ray high and low transitions in the nearby (1-2 kpc) tMSPs PSR J1023+0038 \citep{Stappers14,Patruno14} and XSS J12270-4859 \citep{Bassa14,Roy15}.  Similar tMSP transitions have not yet been used to identify tMSPs at greater distances, where radio pulsar searches are more difficult; the M28 tMSP was identified when it showed a uniquely bright ($\sim 10^{36}~{\rm erg~s}^{-1}$) accretion-powered X-ray outburst, and then returned to a radio pulsar state \citep{Papitto13}.

In addition to the three confirmed tMSPs discussed above, there are a handful of candidate tMSPs discovered through multi-wavelength studies. These tMSP candidates include 1FGL J1417.7-4407, a gamma-ray source with an MSP counterpart  and spectroscopic signs of mass transfer \citep{Strader15,Camilo16}; 1RXS J154439.4-112820 (3FGL J1544.6-1125), a gamma-ray and X-ray binary showing similar X-ray mode-switching behavior as seen in confirmed tMSPs \citep{Bogdanov15a,Bogdanov16b,Britt17}; and 3FGL J0427.9-6704, an eclipsing low-mass X-ray and gamma-ray binary with rapid X-ray variability \citep{strader16}.

Redbacks are MSPs in which the companion is a hydrogen-rich star showing evidence of mass loss, with mass $\gsim$ 0.2--0.5 M$_\odot$ \citep{Roberts13}. In comparison, black widows are MSPs with mass-losing companions with mass $\ll$ 0.1 M$_\odot$ \citep{Fruchter88,Roberts13}. All three of the confirmed tMSPs so far are redback systems \citep{Archibald13,Papitto13,Roy15}. This has motivated a search for tMSP systems among redbacks and black widow pulsars. Numerous black widows and redbacks have high energy counterparts in X-rays and gamma rays \citep[e.g.][]{Roberts13,Linares14b,Gentile14,Roberts15,Ray12}. In X-rays, redbacks are typically brighter (L$_X \sim 10^{31}$ -- $10^{32}$ erg s$^{-1}$) and show a harder non-thermal X-ray spectrum in comparison with black widows, where the nonthermal X-rays are fainter (L$_X \sim 10^{29}$ -- $10^{31}$ erg s$^{-1}$) and sometimes dominated by thermal X-rays from the MSP surface \citep{Gentile14,Roberts15}. It is thought that nonthermal X-rays in these systems mostly come from an intrabinary shock between the pulsar wind and material flowing from the companion \citep{Bogdanov05,Romani16}.

\subsection{Terzan 5}
Terzan 5 is a massive ($\sim 10^6$ M$_\odot$) globular cluster located $\sim$5.9 kpc away in the Galactic bulge \citep{Valenti07,Lanzoni10}. It contains one of the largest populations of X-ray sources \citep[$\gsim50$;][]{Heinke06b} and the highest number of radio MSPs \citep{Prager17} of any  Galactic globular cluster. Such a high population of X-ray and MSP binaries in Terzan 5, compared to other globular clusters, is attributed to dynamics. The formation rate of X-ray binaries per unit mass in globular clusters is roughly two orders of magnitude higher than that in the Galactic field \citep{Katz75}.  It has been shown that this is due to the higher stellar density in these clusters, which leads to the formation of X-ray binaries through dynamical encounters \citep{Verbunt87,Heinke03d,Pooley03,Jordan04,Bahramian13}, as opposed to formation through isolated binary evolution, which is thought to be the dominant channel in the Galactic field. 

Terzan 5 CX1 (CXOGlb J174804.5-244641) is a variable hard X-ray source in the core of Terzan 5. It was first identified in \chandra\ observations in 2003, when it was the brightest source in the cluster \citep{Heinke06b}. The nature of this source is not yet fully understood. In this paper, we present analysis of all the available \chandra\ data on this source. We also identify a radio continuum counterpart, using data taken as part of the MAVERIC (Milky-way ATCA and VLA Exploration of Radio-sources in Clusters; \citealt{Tremou18}; Tudor et al., in prep.; Shishkovsky et al.; in prep.) survey.
Based on the X-ray variability, hard X-ray spectrum and the possibly steep spectrum radio counterpart, we suggest a tMSP nature for this object. 

We present the data, reduction methods, and analysis in Section \ref{sec_data}, and our results in Section \ref{sec_results}. Finally, we discuss possible scenarios for the nature of the source in Section \ref{sec_disc}. In our analysis we assume a distance of 5.9 kpc to Terzan 5 \citep{Valenti07}. All uncertainties reported/plotted based on observations presented in this work are 68\% confidence.

\section{Data reduction and analysis}\label{sec_data}
\subsection{X-rays: \chandra}\label{sec_xraydata}
Terzan 5 has been observed with \chandra\ numerous times since 2001, as it contains a large number of X-ray binaries, three of which are transients which have shown outbursts with $L_X>10^{36}~{\rm erg~s}^{-1}$ \citep{Heinke03b,Wijnands05,Strohmayer10Atel1,Bahramian14}. We reduced and analyzed all archival \chandra/ACIS observations of Terzan 5 in which all cluster sources are faint/quiescent (Table~\ref{tab_xray_obs}). This excludes observations taken during the outbursts of transient XRBs within Terzan 5 (Obs. IDs 664, 655, 11051, 12454, 13708). Additionally, we analyzed observations from our 2016 campaign (PI: N. Degenaar). All \chandra\ observations in this study were performed with ACIS-S in faint mode, with Terzan 5 on chip S3, and the cluster core on-axis. 

\begin{deluxetable}{lccc}[b!]
\tablenum{1}
\tablecaption{\chandra\ observations of Terzan 5 used in this study, and observed average count rates for Terzan 5 CX1}
\label{tab_xray_obs}
\tablehead{
\colhead{Obs. ID} & \colhead{Date} & \colhead{Exposure} & \colhead{Avg. rate} \\
\colhead{}		 & \colhead{} 	& \colhead{(ks)} & \colhead{(ct/s)}}
\startdata
03798	&	2003-07-13	&	39				&	1.17$(\pm0.05)\times10^{-2}$	\\
10059	&	2009-07-15	&	36				&	5.9$(\pm1.4)\times10^{-4}$	\\
13225	&	2011-02-17	&	30				&	4.2$(\pm1.4)\times10^{-4}$	\\
13252	&	2011-04-29	&	40				&	9.6$(\pm1.6)\times10^{-4}$	\\
13705	&	2011-09-05	&	14				&	8.6$(\pm2.7)\times10^{-4}$	\\
14339	&	2011-09-08	&	34				&	9.3$(\pm1.8)\times10^{-4}$	\\
13706	&	2012-05-14	&	46				&	6.8$(\pm1.3)\times10^{-4}$	\\
14475	&	2012-09-17	&	30				&	2.9$(\pm1.2)\times10^{-4}$	\\
14476	&	2012-10-28	&	29				&	3.0$(\pm1.2)\times10^{-4}$	\\
14477	&	2013-02-05	&	29				&	5.3$(\pm1.5)\times10^{-4}$	\\
14478	&	2013-07-17	&	29				&	4.4$(\pm1.4)\times10^{-4}$	\\
14625	&	2013-02-22	&	49				&	6.5$(\pm1.2)\times10^{-4}$	\\
15615	&	2013-02-23	&	84				&	4.5$(\pm0.8)\times10^{-4}$	\\
14479	&	2014-07-15	&	29				&	5.7$(\pm1.7)\times10^{-4}$	\\
16638	&	2014-07-17	&	72				&	6.3$(\pm1.0)\times10^{-4}$	\\
15750	&	2014-07-20	&	23				&	7.8$(\pm1.9)\times10^{-4}$	\\
17779	&	2016-07-14	&	69				&	8.8$(\pm0.4)\times10^{-3}$	\\
18881	&	2016-07-15	&	65				&	8.2$(\pm0.4)\times10^{-3}$	\\
\enddata
\tablecomments{Net count rates are in the 0.5-10 keV band. We omitted all observations taken during outburst of transient XRBs within Terzan 5 (Obs. IDs 664, 655, 11051, 12454, 13708).}
\end{deluxetable}

We used \textsc{ciao} 4.9 and \textsc{caldb} 4.7.3 \citep{Fruscione06} for data reprocessing and analysis. We reprocessed all \chandra\ data with \texttt{chandra\_repro} and extracted source and background spectra using \texttt{specextract}. CX1 is located in the crowded core of Terzan 5, $\sim5''$ from the center of the cluster (Fig.~\ref{fig_cx1_img}). Thus we used a circular extraction region for the source with a radius of 1$''$ to avoid contamination from nearby sources, and an annulus with inner and outer radii of 1.5$''$ and 6$''$ for the background, while excluding all detected sources in this region. Finally, we used \textsc{xspec} 12.9.1n \citep{Arnaud96} for spectral analysis.

\subsection{Radio continuum: VLA}
Deep radio continuum observations of Terzan 5 were obtained with the NSF's Karl G.~Jansky Very Large Array (VLA) in 2012 and 2014. The 2012 observations were taken as part of the MAVERIC survey to study radio sources in globular clusters (Tudor et al., in prep.; Shishkovsky et al.; in prep.), while the 2014 data were obtained to study a Terzan 5 pulsar. In addition to these previously unpublished data, we also use shallower VLA observations from 2015 obtained to study an outburst of the Terzan 5 low-mass X-ray binary EXO 1745--248 \citep{TetarenkoA16}.

For the 2012 observations we observed for five hours at C-band on Sep 10--12 (with two 1 GHz basebands centered at 5.0 and 7.5 GHz) and five hours at S-band on Sep 13--14 (with the two 1 GHz basebands centered at 2.5 and 3.5 GHz). The VLA was in BnA configuration. This bandwidth was sampled with 2 MHz-wide channels and was obtained in full Stokes mode. Calibration was carried out using the complex gain calibrator J1751-2524, while 3C48 provided absolute flux calibration and bandpass calibration. 

On 2014 Feb 14, 3.5 hours of S-band data were obtained on Terzan 5 while the VLA was in BnA-to-A move configuration. These data were observed under test program TPUL0001 (PI P.\ Demorest). The bandwidth was 1024 MHz centered at 2.97 GHz, sampled by 0.5 MHz-wide channels and with dual polarization. The complex gain calibrator was J1751-2524, while 3C286 was used for calibrating bandpasses and the absolute flux scale.

Data editing and calibration was carried out using standard routines in AIPS \citep{Greisen03}. For imaging, the bandwidth was split in half in order to minimize frequency-dependent imaging artifacts. Imaging was performed on each frequency chunk separately, using the task \verb|IMAGR| in AIPS with a Briggs robust value of 1. We self-calibrated the 2012 data in order to minimize artifacts around bright sources. Table~\ref{tab_radio_obs} lists the imaging parameters achieved. The 2012 C-band data were observed at low elevation, so the synthesized beam is rather elongated.

The reduction and analysis of the 2015 VLA data is described in \citet{TetarenkoA16}. These data comprise three 45-min X-band blocks observed in B configuration, from 2015 March 19 to April 12. Each block had two 2048 MHz subbands centered at 9.0 and 11.0 GHz. 
The results presented here use all the data for each subband imaged together.

\begin{deluxetable*}{lccccc}
\tablecaption{VLA observations used in this study and imaging parameters achieved in our analysis}
\tablenum{2}
\label{tab_radio_obs}
\tablehead{
\colhead{Date} & \colhead{Freq} & \colhead{Major axis} & \colhead{Minor axis} &  \colhead{Image rms noise} & \colhead{Source flux}\\
\colhead{}	& \colhead{(GHz)} & \colhead{(arcsec)} & \colhead{(arcsec)} & \colhead{($\mu$Jy/beam)} & \colhead{($\mu$Jy)}}
\startdata
2012 Sep 10-14 & 2.6 & 2.35 & 2.05 & 5.7 & \nodata \\
2012 Sep 10-14 & 3.4 & 1.64 & 1.39 & 5.5 & \nodata \\
2012 Sep 10-14 & 5.0 & 2.48 & 0.79 & 4.7 & $16.4\pm4.9$ \\
2012 Sep 10-14 & 7.4 & 1.73 & 0.47 & 3.6 & $< 10.6$ \\
2014 Feb 14    & 2.7 & 1.18 & 0.59 & 5.0 & $17.7\pm5.7$ \\
2014 Feb 14    & 3.2 & 1.00 & 0.50 & 5.5 & $22.0\pm5.9$ \\
2015 Mar/Apr   & 9.0 & 1.60 & 0.80 & 5.0 & $<14.9$ \\
2015 Mar/Apr   & 11.0 & 1.28 & 0.66 & 5.8 & $<17.5$ \\
\enddata
\tablecomments{Major axis and minor axis are FWHM values for the synthesized beam. We were not able to constrain the source flux in S band data from 2012 as the source appears blended with the bright nearby pulsar Terzan 5 M.  Source flux upper limits in this table are 3-$\sigma$.The 2015 observations were taken in portions on March 19th and 24th and April 12th.}
\end{deluxetable*}

\section{Results}\label{sec_results}
\subsection{X-ray variability and spectral analysis}\label{sec_xrayres}
Comparing \chandra\ observations of CX1, the source shows clear variations in brightness between observations (Table~\ref{tab_xray_obs}, Fig.~\ref{fig_lngterm_lc}). While in 2003 and 2016, CX1 has an X-ray luminosity of L$_X\gsim 10^{33}$ \ergs, in all observations between 2009 and 2014 it shows an average X-ray luminosity of $\sim 1.3\times10^{32}$ \ergs. This indicates that CX1 shows distinct bright and faint ``states''. As we discuss in Section \ref{sec_disc}, these states resemble those seen in tMSP systems.

Besides changes in brightness on month/year timescales, CX1 also shows variability by a factor of a few within observations on timescales of minutes/hours, clearly visible in the 2003 and 2016 observations, when the source was in the bright state (Fig.~\ref{fig_perobs_lc}).

There is no evidence for varability in 
the observations taken during the faint state, using a $\chi^2$ test. However, this apparent lack of variability could be due to our low signal-to-noise ratio in the faint-state observations, caused by the large distance and extinction towards Terzan 5 (Terzan 5 is just slightly further away than M 28, the host cluster of the tMSP IGR J18245-2452, but it suffers almost 10 times more extinction). We test this possibility by artificially degrading CX 1's light curves from observations taken during the bright state (2003 and 2016) so that the average source count rate in these observations match the average rate estimated in the faint state. This degradation increases the uncertainties estimated on counts (and subsequently rates) significantly\footnote{Due to low number of counts, all light curve uncertainties in this study are 1-$\sigma$ estimates based on a Poisson distribution \citep{Gehrels86}.}, enough to make the variations undetectable in a $\chi^2$ test.

\begin{figure*}
\includegraphics[scale=0.95]{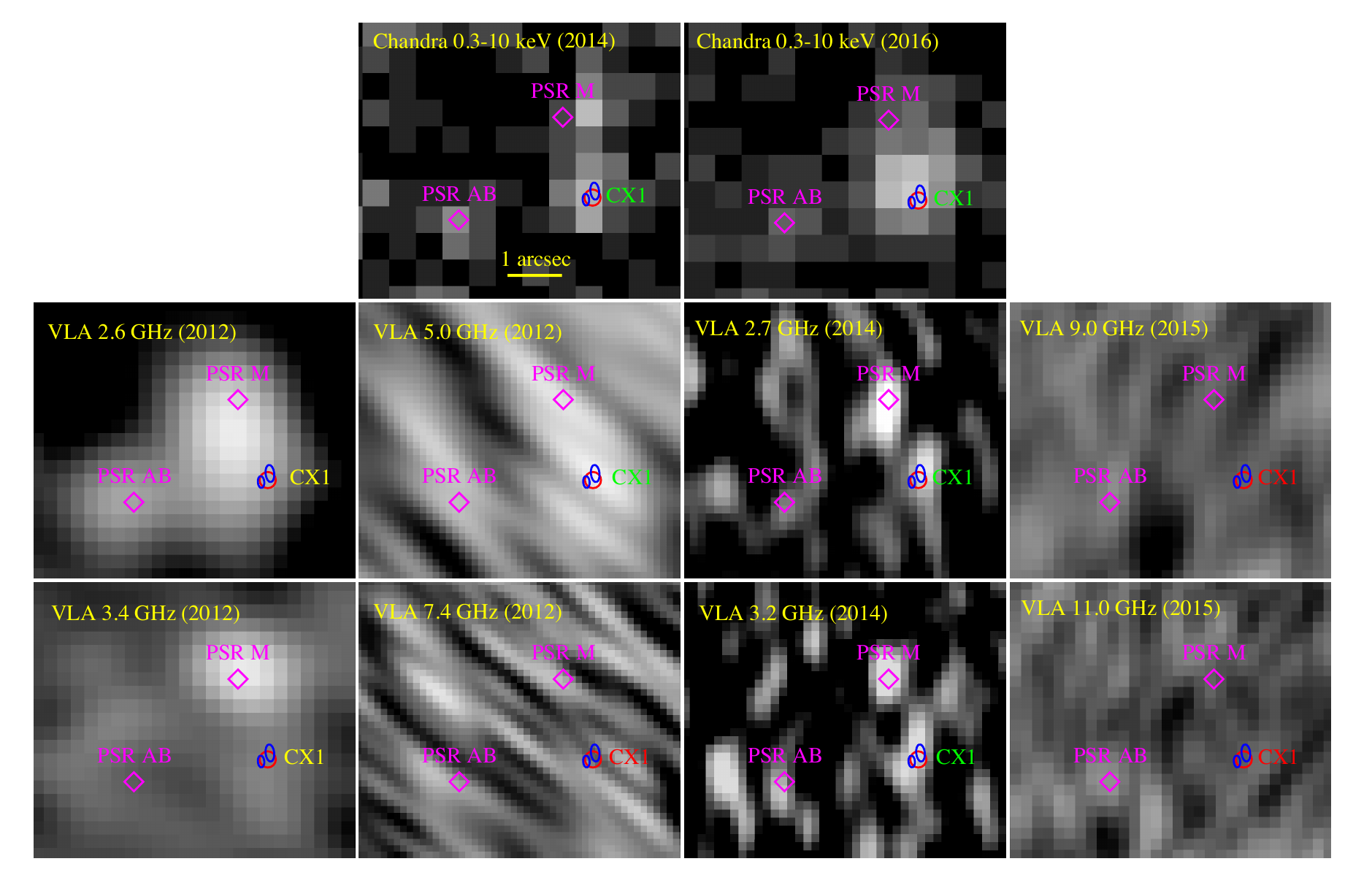}
\caption{\chandra\ (top row) and VLA (middle and bottom rows) observations of Terzan 5,
showing the detection of the source in both X-rays and some radio continuum images.
The red circle shows CX1's localization from X-rays (with an uncertainty of $\approx$ 0\arcsec.15), and blue ellipses represent localizations from the 2014 VLA data (the ellipse to the north estimated based on the 2.7 GHz band and the one to the east based on 3.2 GHz). PSRs Terzan 5 M and Terzan 5 AB are shown with magenta diamonds for comparison. CX1 shows clear variations in X-rays (\chandra\ observations shown here have similar exposure times, 72 ks in 2014 and 69 ks in 2016). CX1 is denoted with green text in images in which it is detected, red where it is not detected and yellow where blended/indeterminate. In the 2012 2.6 and 3.4 GHz observations, CX1 may be present, but no conclusive detection is possible due to blending with nearby sources. Note that colorbars are different for each image.}
\label{fig_cx1_img}
\end{figure*}

\begin{figure}
\begin{center}
\includegraphics[scale=0.5]{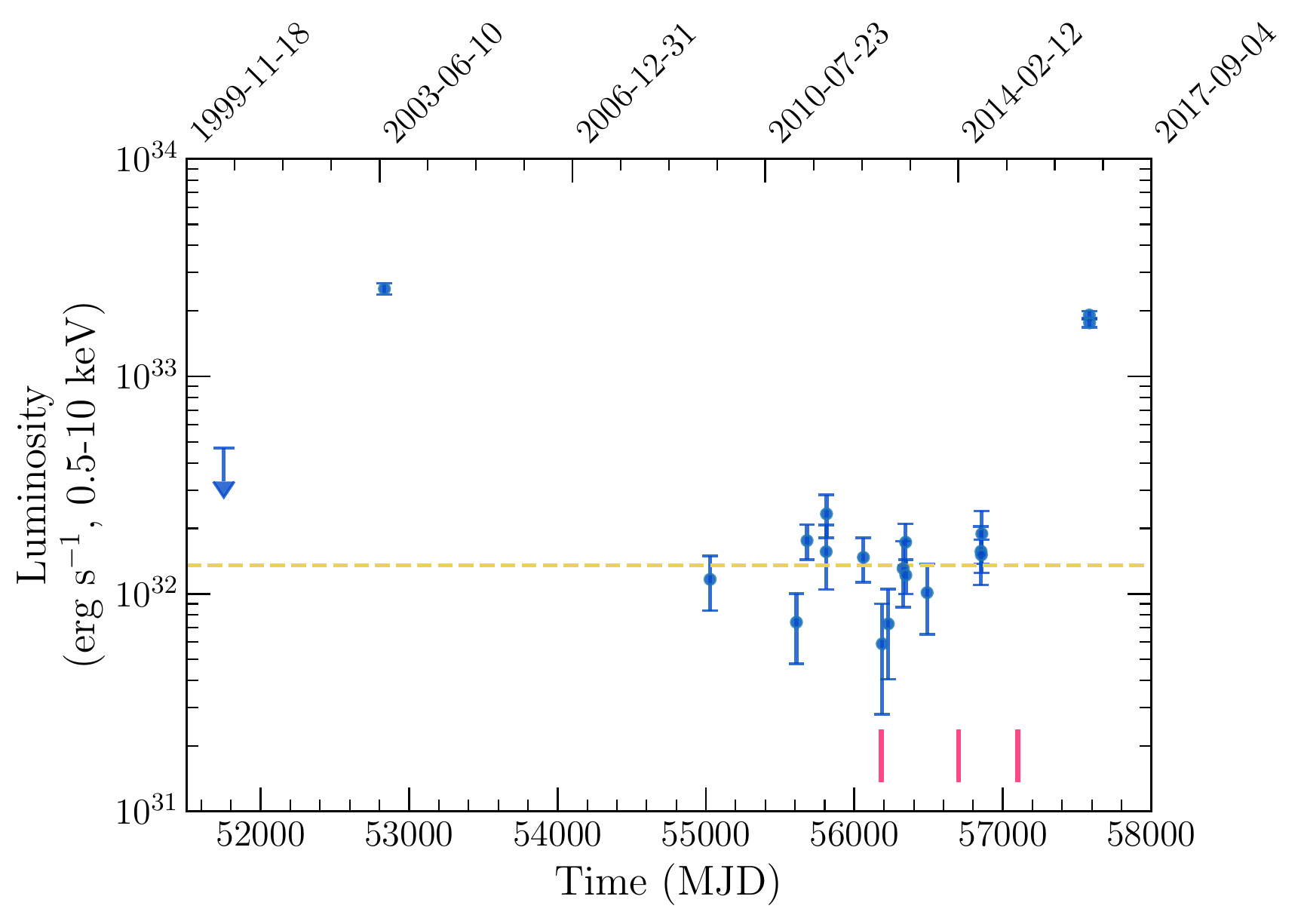}
\caption{\chandra\ 0.5-10 keV light curve of Terzan 5 CX1. The horizontal dashed yellow line represents the average luminosity between 2009 and 2014. The magenta vertical lines indicate the times of our VLA observations. The upper limit arrow represents an estimate from the 2000 \chandra\ observation of Terzan 5 (taken during an outburst from EXO 1745-248), which suggests the source was in a faint state during that observation. This is based on the upper limit reported in \citet{Heinke06a}, after correcting for distance and hydrogen column density differences.}
\label{fig_lngterm_lc}
\end{center}
\end{figure}

\begin{figure*}
\begin{center}
\includegraphics[scale=0.6]{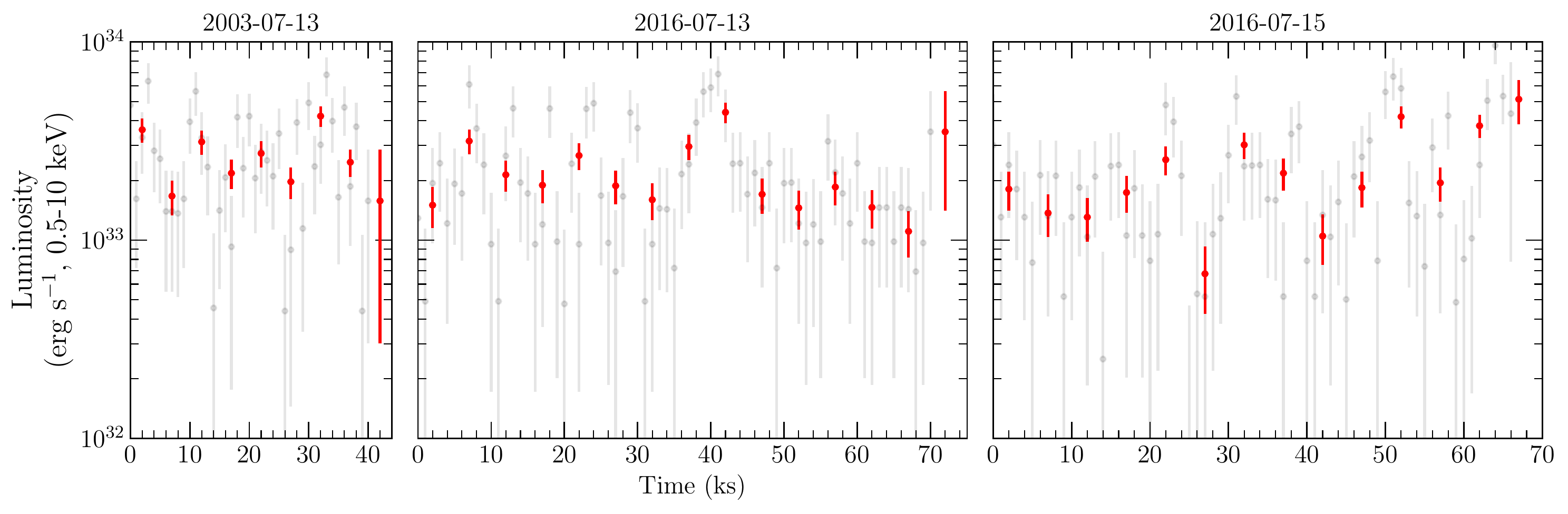}
\caption{\chandra\ lightcurve of CX1 from observations during the bright state. Gray and red show different choices of binning (1 and 5 ks, respectively). There is clear evidence for variability on timescale of minutes/hours (KS test indicates a probability of constancy of $10^{-16}$ for these observations), characteristic of tMSPs.  To convert count rates to luminosities, data points in each light curve were multiplied by the ratio of unabsorbed flux (as obtained from spectroscopy, see \S\ref{sec_xrayres}) to average count rate in the relevant observation.}
\label{fig_perobs_lc}
\end{center}
\end{figure*}

We used \textsc{xspec} 12.9.1 \citep{Arnaud96} for spectral analysis. We assumed \citet{Wilms00} elemental abundances, and \citet{Verner96} photoelectric absorption cross-sections. While spectra taken during the bright state (in 2003 and 2016) each have more than 400 counts, the spectra from observations during the faint state (2009 -- 2014) have only a few counts 
($\sim$ 10 to 50). Since we have no evidence of variation between these observations (Table~\ref{tab_xray_obs}, Fig.~\ref{fig_lngterm_lc}), we combined all spectra from these observations using the \textsc{CIAO} task \textsc{combine\_spec}. We binned all the final spectra (3 during the bright states in 2003 and 2016, and the combined spectrum of observations between 2009 and 2014) by 20 counts per bin, and used chi-square statistics for fitting and goodness-of-fit tests.

We fit each of these four spectra separately with an absorbed power-law (tbabs$\times$pegpwrlw in \textsc{xspec}). As we find no evidence for intrinsic or variable absorption in this system, we set the hydrogen column density (N$_{\text{H}}$) to the value estimated by X-ray spectral fitting for the cluster, $2.6\times10^{22}$ cm$^{-2}$ \citep{Bahramian15}.

\begin{figure}
\begin{center}
\includegraphics[scale=0.35]{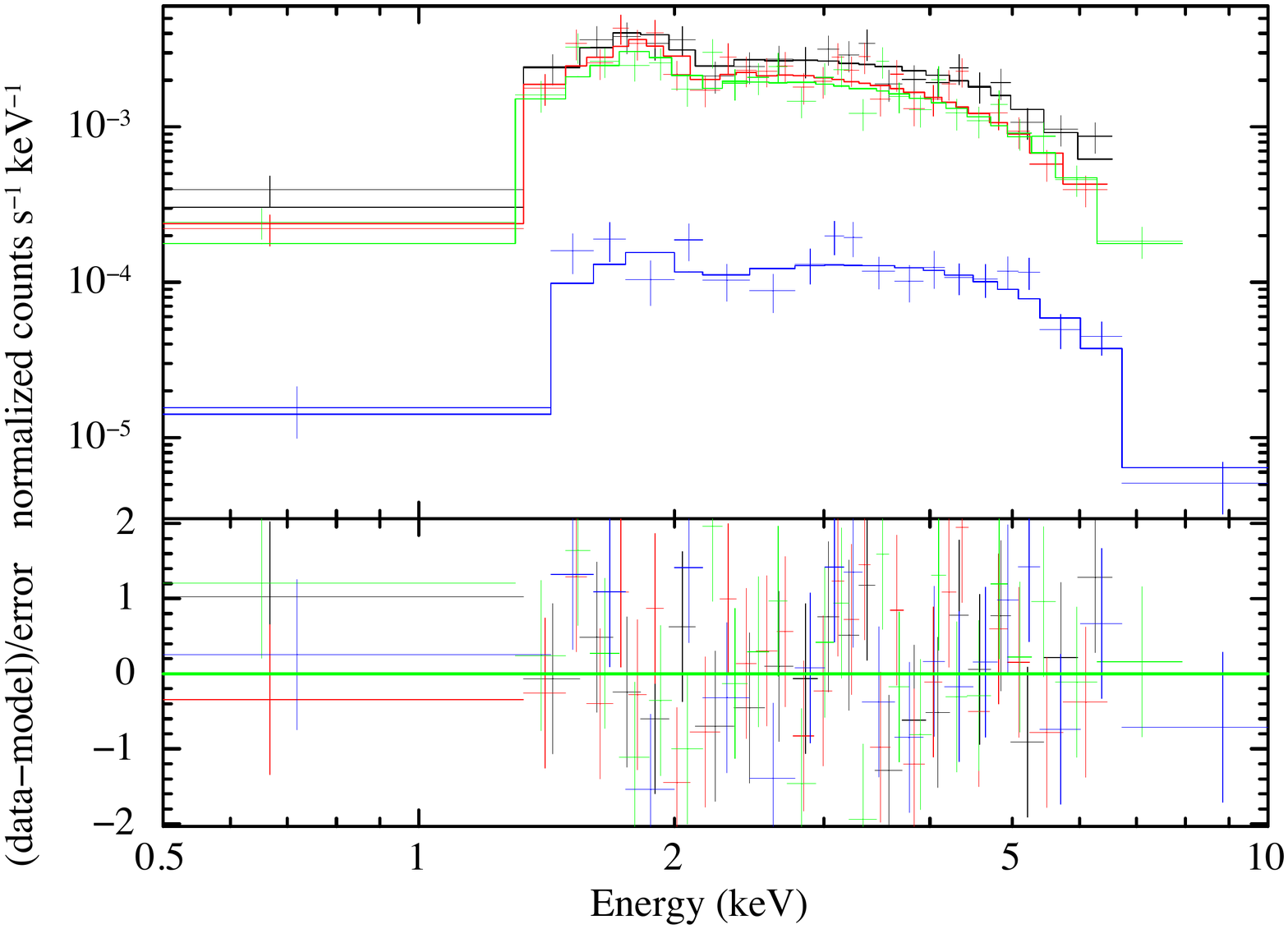}
\caption{Folded \chandra\ X-ray spectra of Terzan 5 CX1 as seen in 2003 (black) and 2016 (red and green), during the bright state, and in 2009 to 2014 (merged spectrum, blue), during the faint state. The fitted model here is an absorbed power law (see \S~\ref{sec_xrayres}). The source shows a harder spectrum during the faint state (see Table \ref{tab_xray_spec}).}
\label{fig_xspec}
\end{center}
\end{figure}

The results of our spectral analysis are tabulated in Table~\ref{tab_xray_spec} and the spectra are plotted in Fig.~\ref{fig_xspec}. A power-law model fits all the spectra well. While all the spectra are hard, there is clear evidence for spectral variability. The spectrum seems hardest in the faint state with a photon index of 0.7$_{-0.1}^{+0.2}$. It is worth noting that the spectra from the bright states in 2003 and 2016 are not completely similar either. In 2003, the spectrum appears comparatively harder (with a photon index of 1.1$\pm0.1$) and the source seems slightly brighter. 

Finally, we note that a precise X-ray position of CX1, tied to a high-quality astrometric frame, is required for multi-wavelength matching of possible counterparts. We start from the coordinates given in the X-ray catalog of \citet{Heinke06b}. To check the absolute astrometry, we note that there are clear X-ray matches with two known millisecond pulsars with accurate positions from radio timing \citep{Prager17}. In addition to the match between CX10 and the pulsar Terzan 5 P noted in \citet{Heinke06b}, the source CX13 appears to be the counterpart of pulsar Terzan 5 ad (analysis of the X-ray counterparts of Terzan 5 millisecond pulsars will be published separately; Bogdanov et al. {\it in prep.}). These two matches show evidence for a consistent offset of about --0.12\arcsec\ in R.A. and +0.29\arcsec\ in Dec. When these offsets are applied to the CX1 position listed in \citet{Heinke06b}, we find an updated J2000 position of (R.A., Dec.) = (17:48:04.578, --24:46:42.24), with an uncertainty of 0.15\arcsec, which we use for the remainder of the paper. A deeper catalog of X-ray sources in Terzan 5, useful for finding additional matches, is beyond the scope of this paper (see Bahramian et al. {\it in prep.}).

\begin{deluxetable}{lccll}
\tablecaption{Results of X-ray Spectral analysis}
\tablenum{3}
\label{tab_xray_spec}
\tablecolumns{4}
\tablehead{ \colhead{Dataset}	&	\colhead{$\Gamma$} &	\colhead{Flux}					&  \colhead{$\chi^2_{\nu}$/d.o.f}	& \colhead{N.H.P}}
\startdata
2003	&	1.1$\pm$0.1	 &6.1$(\pm0.4)\times10^{-13}$&		0.55/20			&	94.4	\\
2009-2014&0.7$_{-0.1}^{+0.2}$&3.3$(\pm0.3)\times10^{-14}$&	1.04/18			&	41.0	\\
2016A	&	1.5$\pm$0.1	 &4.6$(\pm0.2)\times10^{-13}$&		0.83/27			&	70.9	\\
2016B	&	1.4$\pm$0.1	 &4.2$(\pm0.2)\times10^{-13}$&		1.09/24			&	34.9	\\
\enddata
\tablecomments{All uncertainties are 68\% confidence. Unabsorbed flux reported in the 0.5--10 keV band in erg s$^{-1}$ cm$^{-2}$. N.H.P. is XSpec's null hypothesis probability in percent. Dataset 2009-2014 is the combined spectrum from all observations between 2009 and 2014 (see \S~\ref{sec_xraydata}). 2016A spectrum is from observation on 2016-07-14 (obs.ID 17779) and 2016B is from 2016-07-15 (obs.ID 18881).}
\end{deluxetable}

\subsection{Radio Counterpart: spectrum and variability}
CX1 is clearly detected as a radio source in a subset of the VLA observations, but is marginally or not detected in others.

The 2014 S-band data were taken in the most extended configuration of our VLA datasets and have the best angular resolution. CX1 is detected in both subbands, with flux densities of $17.7\pm5.7~\mu$Jy (2.7 GHz) and $22.0\pm5.9~\mu$Jy (3.2 GHz). At 2.7 GHz the J2000 position of this radio source is (R.A., Dec.) = (17:48:04.575$\pm0.08\arcsec$, --24:46:42.12$\pm0.16\arcsec$), matching the X-ray position of CX1 to within 0.13\arcsec. At 
3.2 GHz the radio source is at (R.A., Dec.) = (17:48:04.587$\pm0.06\arcsec$, --24:46:42.28$\pm0.12\arcsec$), also about 0.13\arcsec\ from the position of CX1, even though the 3.2 GHz position is possibly affected by a faint source located southeast of the main source (this faint source is not detected at 2.7 GHz). In any case, the excellent astrometric match between CX1 and the radio source in both subbands (separation of less than 0.15\arcsec, or 1$\sigma$, see Fig.~\ref{fig_cx1_img}) strongly suggests it is the counterpart to CX1.

In the 2012 S-band VLA data (with subbands centered at 2.6 and 3.4 GHz after RFI flagging) there is marginal evidence of a source at the position of CX1, but it is blended with the bright nearby pulsar Terzan 5 M, and we were unable to obtain reliable flux densities, even when simultaneously fitting for the flux density of the pulsar. 

In the 2012 C band data, we detected a radio source at the position of CX1 at 5.0 GHz. Its flux density is $16.4\pm4.9~\mu$Jy. The source was not detected at 7.4 GHz, with a 3-$\sigma$ upper limit of $10.6~\mu$Jy. 

None of the individual-epoch radio observations alone provide meaningful constraints on the radio spectral index of the source. Under the assumption that the source did not vary during the time period of the VLA observations (all of which were in the X-ray faint state), we can constrain the spectral index by modeling the individual detections as well as the upper limits. For a power law with
$S \propto \nu^{\alpha}$ (where $S$ is the flux density, $\nu$ is the frequency, and $\alpha$ the spectral index), we find $\alpha = -1.05^{+0.48}_{-0.65}$, where the uncertainties represent a 68\% confidence interval. This spectral index does not strongly constrain the nature of the source: it is consistent within $1\sigma$ with the steep negative spectral indices typical of millisecond pulsars \citep[$\alpha \sim$ --1.6 to --1.8;][]{kramer98}, but is also consistent within $2\sigma$ with the flat spectra expected for self-absorbed compact jets ($\alpha\sim0-0.6$).

One concern in these results is the potential contamination from the nearby source in the 2014 data (see Fig.~\ref{fig_cx1_img}). The peak flux density for CX1 from the image is consistent with the fitted value within the uncertainty, so there is no evidence that the flux density measurement is significantly biased. To the extent that it has an effect, it would probably bias  the flux density slightly upwards, which would tend to suggest a flatter spectral index. To quantify this, we took the peak flux density of the central pixel of CX1 (18.5 $\mu$Jy) and re-ran the spectral index analysis. We find a value of $-1.02^{+0.51}_{-0.71}$, identical (within uncertainties) to the value mentioned above. Hence there is no evidence that this issue has an effect on any of our conclusions.

\subsection{Constraints on an Optical counterpart}\label{sec_optical}
Terzan 5 has been observed by \acs\ in 2003, 2013 and 2015 with the F606W and F814W filters. We reduced and analyzed these images from the Hubble Legacy Archive\footnote{\url{http://hla.stsci.edu/}} to search for an optical counterpart. 

We set the absolute astrometry of the \hst\ images by performing source detection on every image using \texttt{daofind} task provided in \textsc{pyraf} 2.16 \citep{STScI2012} and cross matching to the \gaia\ data release 1 source catalog \citep{Gaia2016} using the task \texttt{ccmap}. This provides mean rms values of 0.03 arcsec, 0.009 arcsec, and 0.007 arcsec for 2003, 2013 and 2015 epochs respectively, which we consider representative of the absolute astrometric uncertainty.

\begin{figure*}
\begin{center}
\includegraphics[scale=0.9]{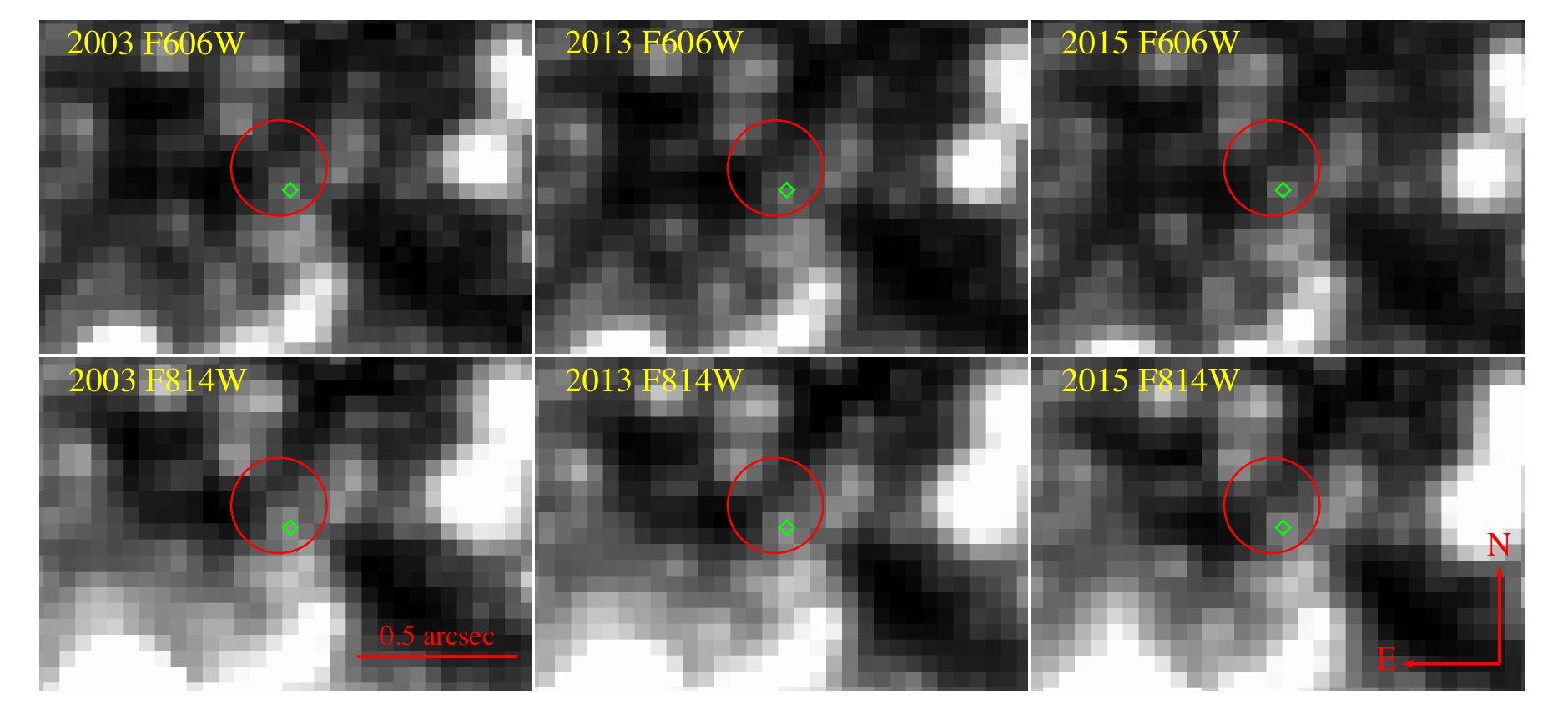}
\caption{\acs\ images of Terzan 5 CX1 (red circle). The source in CX1's error circle (indicated with a green diamond) appears to be a ``normal'' main-sequence star and is unlikely to be the counterpart (see \S~\ref{sec_optical} for details).}
\label{fig_hst}
\end{center}
\end{figure*}

The density of stars in the core of Terzan 5 is extremely high, and there is a source at RA = 17:48:04.575, Dec = --24:46:42.310, formally 0.08\arcsec from CX1 (Fig.~\ref{fig_hst}). Aperture photometry of this source yields mean 
m$_{\text{F606W}}$ and m$_{\text{F814W}}$ magnitudes of $24.4\pm0.5$ and $21.3\pm0.3$ mag (Vega mag), respectively. There is no significant evidence of variation across the \hst\ epochs. Compared to the published color-magnitude diagrams of Terzan 5 in these filters \citep[e.g., see][]{Massari12}, this photometry is consistent with the identification of this star as a likely main sequence star, though the high and variable extinction makes a more precise classification impossible. The optical (and especially infrared) counterparts of many X-ray binaries  in quiescence appear quite like ``normal'' stars, as the optical/infrared emission is completely dominated by the donor star. However, at least two of the HST epochs targeting Terzan 5 are taken around times when CX 1 shows brighter (few $10^{33}$ erg/s) X-ray emission (in 2003 and 2015). Transitional MSPs typically show increases in brightness of 1--2 magnitudes when they change to the bright state \citep{Halpern13,Bassa14}. Thus, the ``normal'' color of this star, as well as its lack of variability over the observed epochs (which are taken around times the source is bright in X-rays), suggests that it is unlikely to be the true counterpart to CX1.

To estimate the probability of a random match with an optical source in the core of Terzan 5, we picked 50000 random coordinates within the core of Terzan 5 and measured the distance to the closest real detection from each (in the deepest \hst\ image). The distribution of these separations appear gaussian with a peak around $0.18$\arcsec and a standard deviation of 0.09\arcsec. This distribution indicates that the probability of finding an optical source  $\leq0.15$\arcsec (uncertainty radius in the X-ray coordinates) is $\sim 35 \%$. This reinforces the conclusion above that the putative optical counterpart might well be an interloper and that the true counterpart may be too faint to be detected in these data.

\section{Discussion}\label{sec_disc}
\subsection{Terzan 5 CX1 as a tMSP}
In X-rays, Terzan 5 CX1 shows two brightness states. The faint state appears to last for  years (considering the many consecutive faint-state observations), and the few observations of the bright state suggest a duty cycle of $\sim1/6$ -- $1/3$. In the bright state, the source shows an average luminosity of L$_{X}\approx2\times10^{33}$ erg s$^{-1}$, and in the faint state a luminosity of L$_{X}\approx1.3\times10^{32}$ erg s$^{-1}$ in the 0.5--10 keV band. These luminosity levels and their timescales are consistent with the low-level accretion (bright) and pulsar (faint) states observed in the tMSP systems. During the pulsar state, PSR J1023+0038 shows an X-ray luminosity between $9\times10^{31}$ and $1.6\times10^{32}$ erg s$^{-1}$ in the 0.5--10 kev band \citep{Archibald10,Bogdanov11,Linares14b}. In the same state, XSS J12270-4859 shows L$_X\approx1.6\times10^{32}$ \citep{Bogdanov14,Papitto15} and IGR J18245-2452 shows L$_X\approx2.2\times10^{32}$ \citep{Linares14a}. These luminosities are typical of redback rotation-powered pulsars \citep{Linares14b,Roberts15,Gentile14}.

In the low-level accretion state, tMSPs tend to show significant variability, and occasional flares \citep[e.g.,][]{Tendulkar14,Takata14,Papitto15}. They can also show two dominant ``modes'' of brightness (labeled active and passive modes), which can be different by a factor of a few. These modes can last hours and are distinct enough that the distribution of X-ray brightness appears as a bimodal distribution \citep[e.g.,][]{deMartino13,Linares14a,Bogdanov15b}.

\begin{figure*}
\begin{center}
\includegraphics[scale=0.47]{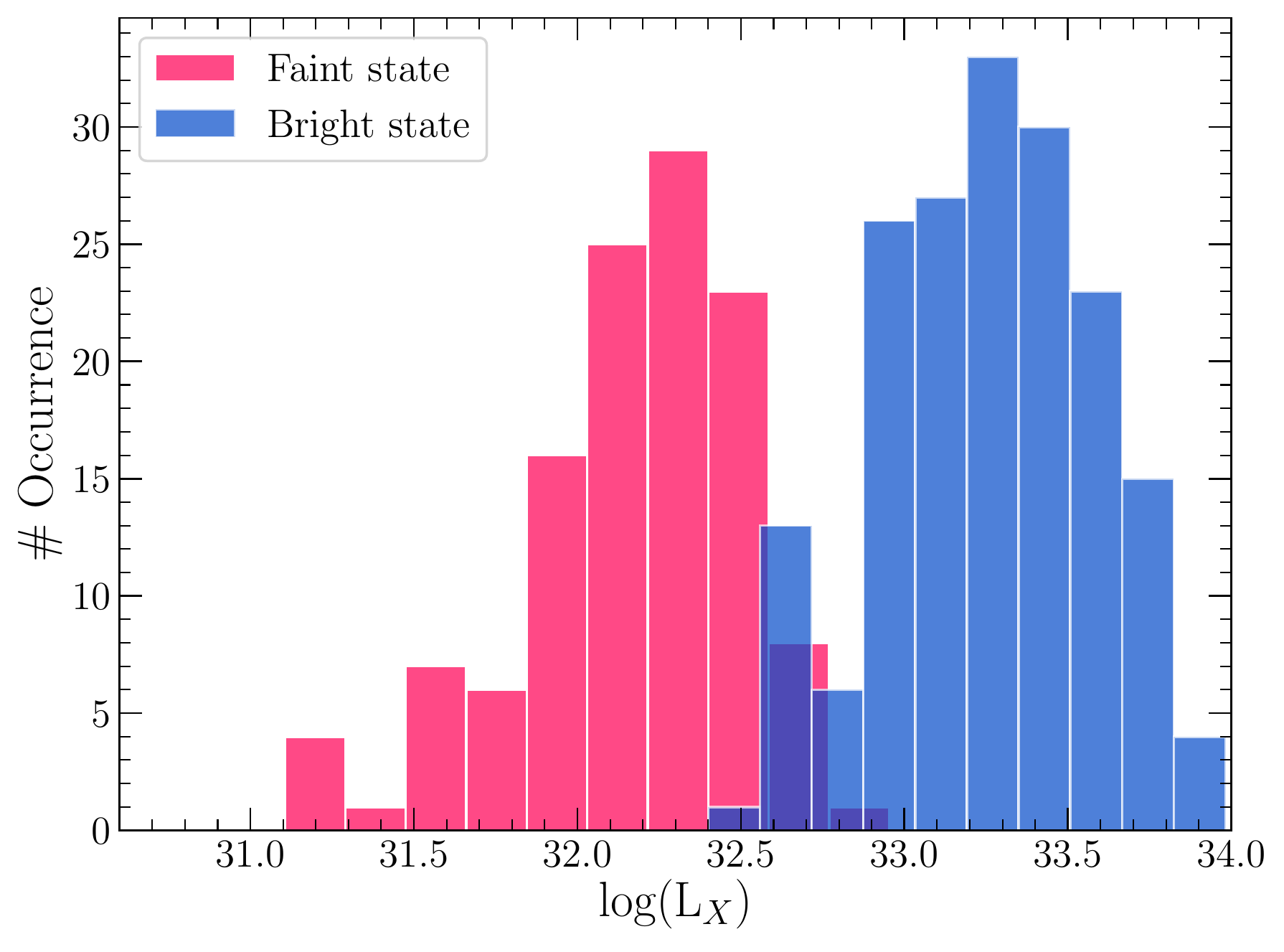}
\includegraphics[scale=0.47]{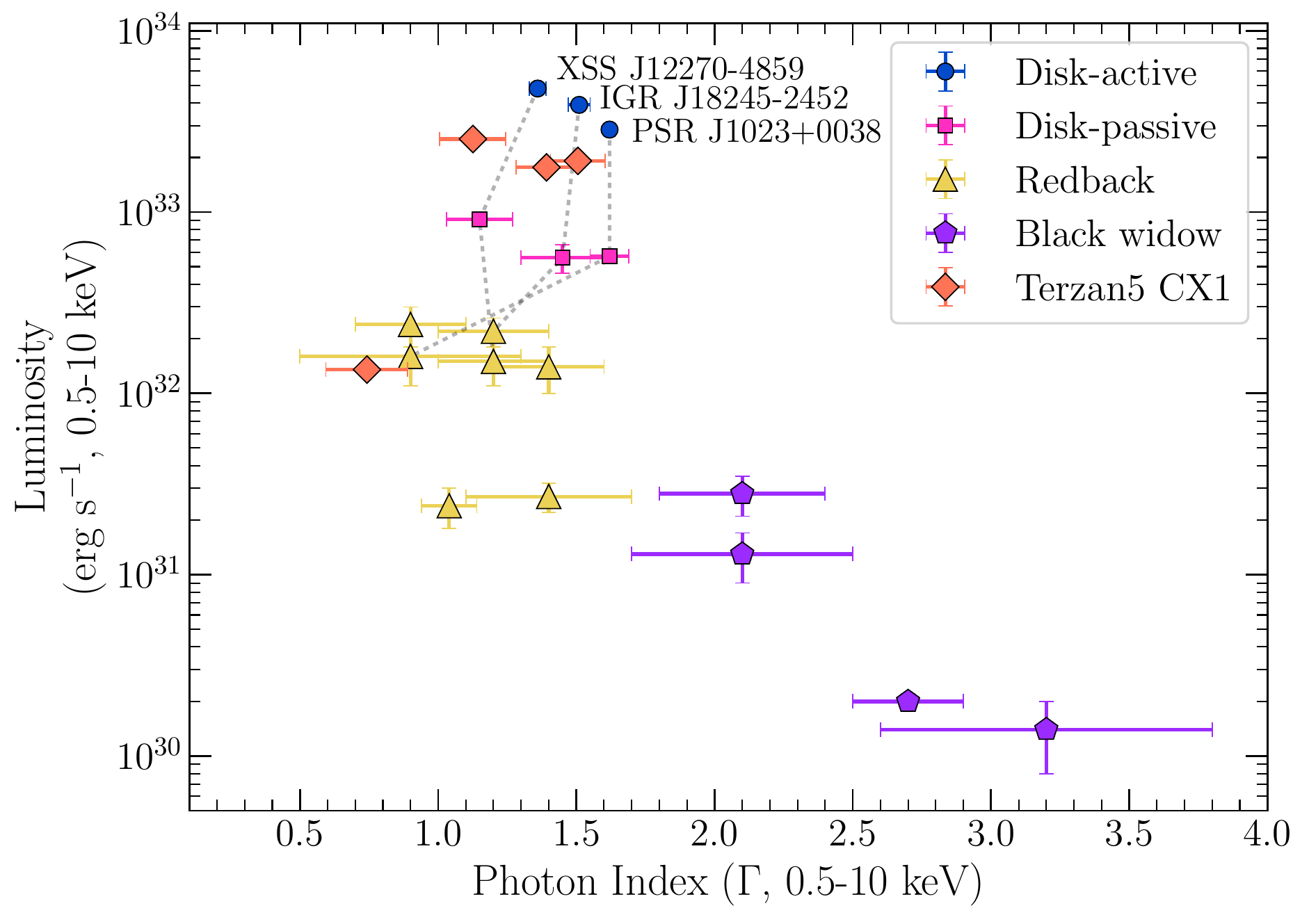}
\caption{{\bf Left}: Distribution of brightness in Terzan 5 CX1. The blue data set represents the distribution based on observations during the bright state (similar to the low-level accretion state in tMSPs), while the pink dataset shows the distribution during the faint state (similar to the pulsar state in tMSPs). There is no clear bimodality visible in the bright state indicating the presence of high and low modes as have been seen in tMSPs. Luminosity is in the 0.5-10 keV band in erg s$^{-1}$. {\bf Right}: Hardness-luminosity diagram of redbacks (yellow triangles), black widows (purple pentagons) and low-level accretion states in tMSPs with modes separated (active mode represented by blue circle and passive mode by pink squares). The dashed lines shows the evolution of tMSPs between the accretion state and the pulsar state. Terzan 5 CX1 is shown by orange diamonds. In the bright state, Terzan 5 CX1 behavior is consistent with average tMSP luminosity and hardness in the accretion state. In the faint state, it is very similar (slightly harder) to redbacks. Plot adapted from \citet{Linares14b}. Data for PSR J2129-0429 is from \citet{Roberts15}. Data for black widows and PSR J2215+5135 is from \citet{Gentile14}.}
\label{fig_tmsp_comparison}
\end{center}
\end{figure*}

Terzan 5 CX1 shows clear variability during the bright state. A KS test indicates a probability of constancy of $10^{-16}$ for observations during the bright state. Some of these variations appear to be flares and at times the system stays at one brightness level for a few ksec (Fig.~\ref{fig_perobs_lc}), not unlike modes observed in tMSPs. However, the X-ray brightness does not show a clear bimodal distribution (Fig.~\ref{fig_tmsp_comparison}, left). However, there is a tentative peak in the histogram at $\log(L_X)\sim32.5$ erg s$^{-1}$, hinting at a possible bimodal distribution. The absence of a clear bimodal flux distribution in this case could be due to sensitivity, as for PSR J1023+0038 - which is significantly closer and less extincted than Terzan 5 - \citet{Bogdanov15b} find the longest low mode to be 1.9 ks, and the longest high mode 6.9 ks, while most of the periods in a single mode are well under 1 ks (see their figure 6). This is shorter than the timescales we are sensitive to in \chandra\ observations of Terzan 5, see Fig.~\ref{fig_perobs_lc}. To test this further, we downloaded the PSR J1023+0038 \emph{XMM-Newton/EPIC} light curve from obsID 0720030101, which shows numerous low and high modes (reported in \citealt{Bogdanov15b}), and rebinned the light curve to 1 ks per bin (chosen as a timescale similar to our sensitivity levels in \chandra\ light curve of Terzan 5 CX 1). Doing so completely removes the bimodal distribution. However, while PSR J1023+0038 and XSS J12270-4859 tend to show short low modes, the low mode has been observed to last for $\sim$ 10 ks in IGR J18245-2452 \citep{Linares14a}. Thus, we recognize that a key aspect of the X-ray phenomenology of known tMSPs, a bimodal count rate in the disk state, could only be detected in Ter5 CX1 if it follows the longer-duration pattern seen in IGR J18245-2452.

tMSPs (in the low-level accretion state and the pulsar state) and redback pulsars tend to show hard X-ray spectra, generally well-described by a power-law model with photon index $\leq 1.6$ \citep[e.g.,][]{Linares14b,Roberts15}. On average, the X-ray spectra of tMSPs are strikingly similar between the different states (as opposed to NS LMXBs for instance, which can change significantly in slope over 1 magnitude in X-ray luminosity, see e.g., \citealt{Wijnands15}). Terzan 5 CX1's spectra seem to follow a similar trend as the tMSPs  (Fig.~\ref{fig_tmsp_comparison}, right). While the bright state in this system cannot be clearly broken into ``high'' and ``low'' modes, the hardness and luminosity are consistent with the average luminosity and hardness of tMSPs in the low-level accretion state, and the luminosity and hardness of the faint state of this system are consistent with those of redbacks (including tMSPs).

The 2012 VLA data were taken within a week of a \chandra\ observation finding CX1 in its faint state, strongly indicating it was X-ray faint during those observations. The 2014 VLA data were 5 months from the nearest \chandra\ observation, but as all \chandra\ observations within 2 years of this observation found CX1 to be X-ray faint, we reasonably assume these VLA data were also taken in the faint state.  In the tMSP interpretation, the X-ray faint state implies a radio pulsar state, which is indeed slightly preferred by our limited constraints on the radio spectral index.  The 2015 VLA data were observed between a 2014 \chandra\ observation in the X-ray faint state, and a 2016 \chandra\ observation in the bright state, so we do not know CX1's X-ray state in 2015.  The upper limits from the 2015 VLA data are not constraining for either hypothesis.  Assuming that, in the bright state, CX1 has a ratio of radio to X-ray flux comparable to that of the tMSP PSR J1023+0023 (Deller et al. 2015), then the predicted mean radio flux density in the 2015 X band observations would be $\sim$5 $\mu$Jy, too faint to be detected in the 2015 VLA data (which had a stacked rms noise of 3.8 $\mu$Jy/beam.).

\subsection{Other possibilities for the nature of Terzan 5 CX1}
Terzan 5 CX1 is located in the core of Terzan 5, $\sim$ 5$''$ from the center of the cluster. This makes the likelihood of a chance alignment extremely small: it is almost certainly associated with the cluster. This dramatically reduces the possibility of it being an AGN or a foreground flare star. The absence of a bright optical counterpart, and the absorption of its X-ray spectrum, also argue against a foreground flare star nature.

CX1's X-ray luminosity in the bright state is also similar to that of the brightest cataclysmic variables \citep[CVs, e.g.,][]{Stacey11,Bernardini12,Pretorius14}. Many CVs show detectable variability in X-rays and a hard X-ray spectrum. However, in the radio, CX1's persistent emission is brighter than that of the brightest CVs. An example is SS Cyg, a CV with similar X-ray luminosity as CX1 during the faint state, which shows bright radio flares up to $\sim 1.4\times10^{27}$ erg s$^{-1}$ at 5 GHz \citep{Russell16,Mooley17}. While these flares show radio luminosities similar to that of CX1, they are very short-lived (lasting $\sim$ minutes), while CX1 appears persistent, as  measured in multiple epochs.

Many CVs show strong Fe K$\alpha$ lines in their X-ray spectra \citep[e.g.,][]{Norton91,deMartino08,Kuulkers10,Byckling10,Mukai17}. We do not find any evidence for such a line in the X-ray spectra of CX 1 (even after merging the spectra). However, it is possible that this is due to poor statistics in the X-ray spectra or full ionization of Fe in the emitting region.

The X-ray spectrum of CX 1 is similar to many LMXBs containing black holes or neutron stars. In the X-ray binary scenario for CX 1, the bright X-ray detections in 2003 and 2016 could be explained as accretion outbursts. However, most LMXBs  spend most of their outburst at X-ray luminosities above $10^{35}$ erg s$^{-1}$. This would mean that both 2003 and 2016 observations were taken during the early rise or late decay of potential outbursts from CX 1, which seems unlikely. 

One class of X-ray transients known as very faint X-ray transients (VFXTs) undergo faint outbursts with X-ray luminosity  peaking between $\sim10^{34}$ and $10^{36}$ erg s$^{-1}$ during the outburst. Many such transients have been found in surveys of the Galactic center \citep{Muno05b,Sakano05,Wijnands06,Degenaar09}, and a few are found in globular clusters \citep[e.g.,][]{Heinke09b,Arnason15}. While most of these systems show short outbursts (timescale of hours to days), there are some that show quasi-persistent outbursts \citep[e.g.,][]{ArmasPadilla13a}. The nature of these systems and their low accretion rates are not fully understood, but they are likely to be a somewhat heterogeneous population. It has been suggested that VFXTs might be `period-gap' systems \citep{Maccarone13} or that some of them might be unidentified tMSPs \citep{Heinke14}, and recently \citet{Hofmann18} identified a new transient intermediate polar system as a VFXT. The behavior of CX1 in X-rays is similar to some VFXTs.

It is possible that black hole LMXBs in globular clusters show different outburst behavior than field systems due to their dynamical formation, including systems that show only faint short outbursts \citep{Knevitt14}. This is certainly a fascinating possibility for CX1. However, confirmed accreting neutron stars substantially outnumber even candidate accreting black holes in globular clusters \citep[e.g.][table 5\footnote{\url{https://bersavosh.github.io/research/gc_lmxbs.html}}]{Bahramian14}, and the large MSP population of Ter 5 means that it has hosted many neutron star LMXBs, so in the absence of other constraints, a neutron star accretor should be considered more likely. 

\section{Conclusions}
We conclude that all available evidence is consistent with the hypothesis that Terzan 5 CX1 is a tMSP that was in a low-level accretion state in 2003, 
transitioning to a pulsar state sometime before 2009. The high density of observations between 2009 and 2014 suggest it was largely or entirely in the pulsar state during this time. By 2016, it had returned to a low-level accretion state (Fig.~\ref{fig_lngterm_lc}).

Extending the baseline further, early \chandra\ observations, obtained during the outburst of the low-mass X-ray binary EXO 1745-248 \citep{Heinke03b}, provide a $3\sigma$ upper limit of L$_X < 4.7\times10^{32}$ erg s$^{-1}$. This suggests that in 2000 the system may have been in the fainter pulsar state, indicating the possibility of an additional transition between 2000 and 2003.

Given the extensive radio pulsar timing dataset available for Terzan 5, the best way to confirm CX1 as a tMSP would be to detect the source as a radio pulsar using data obtained during a putative pulsar phase (e.g., searching the data obtained between 2009 and 2014). Such searches are ongoing, and new pulsars are being detected in older Terzan 5 timing observations using refined search techniques \citep[e.g.,][]{Cadelano18} \footnote{Note that the new MSPs discovered by \citet{Cadelano18} are all isolated, and thus unlikely to be associated with Terzan 5 CX1.}. It also suggests that deep, well-separated epochs of X-ray data should be searched for other globular clusters, and could reveal the presence of additional tMSPs.

\section*{Acknowledgments}
We gratefully acknowledge support from NSF grant AST-1308124. JS was partially supported by the Packard Foundation. AJT, COH and GRS are supported by NSERC Discovery Grants, and COH also by a Discovery Accelerator Supplement. AJT is supported by a Natural Sciences and Engineering Research Council of Canada (NSERC) Post-Graduate Doctoral Scholarship (PGSD2-490318-2016). JCAM-J is the recipient of an Australian Research Council Future Fellowship (FT140101082). ND is supported by a Vidi grant awarded by the Netherlands Organization for Scientific Research (NWO). ET acknowledges financial support from the UnivEarthS Labex program of Sorbonne Paris Cit\'e (ANR-10-LABX-0023 and ANR-11-IDEX-0005-02). The National Radio Astronomy Observatory is a facility of the National Science Foundation operated under cooperative agreement by Associated Universities, Inc. 

The scientific results reported in this article are based on observations made by the \chandra\ X-ray Observatory, the Karl G.~Jansky Very Large Array, archival data obtained from \chandra, observations made with the NASA/ESA Hubble Space Telescope and obtained from the Hubble Legacy Archive, and data from the European Space Agency (ESA) mission \gaia. We acknowledge extensive use of NASA's Astrophysics Data System, Arxiv, and Vizier.

\vspace{5mm}
\facilities{Chandra (ACIS), VLA, HST (ACS)}
\software{
	AIPS \citep{Greisen03},
	Astropy \citep{Robitaille13}, 
    CIAO \citep{Fruscione06},
    HEASOFT \citep{HEASARC14},
    Matplotlib \citep{Hunter07},
    PyRAF \citep{STScI2012},
    SAOImage DS9 \citep{Smithsonian2000},
    Xspec \citep{Arnaud96}
}

\bibliography{full_bibliography.bib}
\listofchanges
\end{document}